\documentclass[preprint]{aastex63}

\usepackage{color}
\usepackage{natbib}
\usepackage{amsmath}
\usepackage{nicefrac}
\tightenlines

\newcommand \AU	        {\,{\rm AU}}
\newcommand \bahat      {\hat{\bf a}}
\newcommand \bB         {{\bf B}}

\newcommand \bbhat      {\hat{\bf b}}

\newcommand \bE         {{\bf E}}
\newcommand \behat      {\hat{\bf e}}

\newcommand \bJ         {{\bf J}}

\newcommand \br         {{\bf r}}

\renewcommand \bv       {{\bf v}}
\newcommand \bvdrift    {{\bf v}_{\rm drift}}
\newcommand \bxhat      {\hat{\bf x}}
\newcommand \byhat      {\hat{\bf y}}
\newcommand \bzhat      {\hat{\bf z}}

\newcommand \bz         {{\bf z}}
\newcommand \beq        {\begin{equation}}
\newcommand \beqa	{\begin{eqnarray}}

\newcommand \calP       {{\cal P}}

\newcommand \cm         {\,{\rm cm}}

\newcommand \eeq	{\end{equation}}
\newcommand \eeqa	{\end{eqnarray}}

\newcommand \erg	{\,{\rm erg}}

\newcommand \GHz        {\,{\rm GHz}}
\newcommand \gtsim	{\gtrsim}		 

\newcommand \Ha 	{{\rm H}}

\newcommand \K  	{\,{\rm K}}

\newcommand \Lsol	{L_{\odot}}
\newcommand \ltsim	{\lesssim}		 

\newcommand \MJy        {\,{\rm MJy}}

\newcommand \Msol	{M_{\odot}}

\newcommand \nH         {n_{\rm H}}

\newcommand \NH         {N_{\rm H}}

\newcommand \pc  	{\,{\rm pc}}

\newcommand \sr  	{\,{\rm sr}}

\newcommand \mm         {\,{\rm mm}}

\newcommand \falign     {f_{\rm align}}

\newcommand{\btdnote}[1]{}

\countdef\decade=200
\advance\decade by \year
\countdef\hours=201
\advance\hours by \time
\divide\hours by 60
\countdef\mins=202
\advance\mins by \hours
\multiply\mins by 60
\multiply\hours by 100
\countdef\miltime=203
\advance\miltime by \hours
\advance\miltime by \time
\advance\miltime by -\mins

\begin{document}

\title{%
        {\bf On Far-Infrared and Submm Circular Polarization}
	}

\author[0000-0002-0846-936X]{B.~T.~Draine}
\affiliation{Dept.\ of Astrophysical Sciences,
  Princeton University, Princeton, NJ 08544, USA}

\email{draine@astro.princeton.edu}

\begin{abstract}
Interstellar dust grains are often aligned.
If the
grain alignment direction varies along the line of sight, 
the thermal emission becomes circularly-polarized.
In the diffuse interstellar medium,
the circular polarization at far-infrared
and submm wavelengths is predicted to be very small, and probably
unmeasurable.  However,
circular polarization may reach detectable levels in
infrared dark clouds and protoplanetary disks.  Measurement of circular
polarization could help constrain the structure of the magnetic field
in infrared dark clouds, and may shed light on the mechanisms responsible
for grain alignment in protoplanetary disks.
\end{abstract}

\let\svthefootnote\thefootnote
\let\thefootnote\relax\footnote{\textcopyright 2021.  All rights reserved.}
\let\thefootnote\svthefootnote

\section{Introduction
         \label{sec:intro}}

Since the discovery of starlight polarization over 70 years ago
\citep{Hiltner_1949a,Hall_1949}, 
polarization has become a valuable tool for study of both the physical
properties of interstellar dust and the structure of the interstellar
magnetic field.  Starlight polarization arises because
initially unpolarized starlight becomes linearly polarized as a result
of linear dichroism produced by aligned dust grains
in the interstellar medium (ISM).
While the physics of dust grain alignment is not yet fully understood,
early investigations \citep{Davis+Greenstein_1951} showed how 
spinning dust grains
could become aligned with their shortest axis
parallel to the magnetic field direction.  Subsequent studies
have identified a number of important physical processes that were
initially overlooked
\citep[see the review by][]{Andersson+Lazarian+Vaillancourt_2015}, 
but it remains clear that in the
diffuse ISM the magnetic field establishes the direction
of grain aligment, with the dust grains tending to align 
with their short axes parallel to the local magnetic field.

\citet{van_de_Hulst_1957} noted
that if the magnetic field direction was not uniform,
starlight propagating through the dusty
ISM would become circularly
polarized.
This was further discussed by \citet{Serkowski_1962} and 
\citet{Martin_1972c}.
The birefringence of the dusty ISM is responsible for
converting linear polarization to circular polarization
\citep{Serkowski_1962,Martin_1972c}.
The strength of the resulting circular polarization depends on the
changes in the magnetic field direction and also on the optical properties 
of the dust.

Circular polarization of optical light from the Crab Nebula
was observed by \citet{Martin+Illing+Angel_1972}.
Circular polarization
of starlight was subsequently observed by
\citet{Kemp_1972} and \citet{Kemp+Wolstencroft_1972};
the observed degree of circular polarization, $|V|/I \ltsim 0.04\%$,
was small but measurable.
As had been predicted, the circular polarization $V$ changed sign 
as the wavelength varied from blue to red, 
passing through zero near the wavelength $\sim$$0.55\micron$ 
where the linear polarization peaked \citep{Martin+Angel_1976}.

Because the circular polarization depends on the change in magnetic field \added{direction}
along the line of sight, it can in principle 
be used to study the structure of the Galactic magnetic field.
Data for 36 stars near the Galactic
Plane suggested a systematic bending of the field for Galactic longitudes
$80^\circ \ltsim \ell < 100^\circ$
\citep{Martin+Campbell_1976}.
However, 
these studies do not appear to have been pursued,
presumably because 
sufficiently bright and reddened stars are sparse.
 
In the infrared, circular polarization has been measured for bright
sources in molecular clouds
\citep{Serkowski+Rieke_1973,
       Lonsdale+Dyck+Capps+Wolstencroft_1980,Dyck+Lonsdale_1981}.
Measurements of linear and 
circular polarization were used
to constrain the magnetic field structure in the
Orion molecular cloud OMC-1
\citep{Lee+Draine_1985,Aitken+Hough+Chrysostomou_2006}.

Circular polarization has also been observed in the infrared 
(K$_{\rm s}$ band) in
reflection nebulae 
\citep{Kwon+Tamura+Hough+etal_2014,
       Kwon+Tamura+Hough+etal_2016,
       Kwon+Nakagawa+Tamura+etal_2018}, 
but in this case
scattering is \deleted{thought to be} important
\citep{Fukushima+Yajima+Umemura_2020}.
\added{Scattering can convert linear to circular polarization}, 
making interpretation dependent
on the uncertain scattering geometry.

It was long understood that the nonspherical and aligned grains responsible 
for starlight polarization must emit far-infrared radiation which would
be linearly polarized.
Observations of this polarized emission now allow the
magnetic field direction projected on the sky to be mapped in the general ISM
\citep[see, e.g.,][]{Planck_int_results_xix_2015,Planck_int_results_xxi_2015,
Fissel+Ade+Angile+etal_2016}.
Ground-based observations have provided polarization
maps for high surface-brightness regions at submm frequencies
\citep[e.g.,][]{Dotson+Vaillancourt+Kirby+etal_2010},
and the Statospheric Observatory for Infrared Astronomy (SOFIA) is
providing polarization maps of bright regions in the far-infrared
\citep[e.g., OMC-1:][]{Chuss+Andersson+Bally+etal_2019}.

ALMA observations of mm and submm emission from protoplanetary disks 
find that the radiation is often linearly polarized.  
Scattering may contribute to the polarization
\citep{Kataoka+Muto+Momose+etal_2015}, 
but the observed polarization directions and wavelength dependence 
appear to
indicate that a substantial fraction of
the polarized radiation arises from thermal
emission from aligned dust grains
\citep{Lee+Li+Yang+etal_2021}.

Previous theoretical discussions of circular polarization 
were mainly concerned with
infrared and optical wavelengths where 
initially unpolarized starlight 
becomes polarized as a result of linear dichroism.
In a medium with changing polarization direction, the resulting circular
polarization is small because the linear polarization itself is
typically
only a few \%, and the optical ``phase shift''
\added{(between the two linear polarization modes)} 
produced by the aligned medium
is likewise small.  At far-infrared wavelengths, 
however, the radiation is already
substantially polarized when it is emitted, with linear polarizations
of 20\% or more under favorable conditions
\citep{Planck_2018_XII}.
While absorption optical depths tend to be small at long wavelengths, 
the optical properties of the dust are 
such that phase shift cross sections at submillimeter wavelengths
can be much larger than 
absorption cross sections, raising the possibility that a medium with changing
alignment direction might exhibit measurable levels of circular polarization
at far-infrared or submm wavelengths.

The present paper discusses polarized radiative transfer in a medium with
partially aligned nonspherical grains, including both absorption and
thermal emission. 
We estimate the expected degree of circular polarization
for emission from molecular clouds and protoplanetary disks.
For nearby molecular clouds, the far-infrared circular polarization is
very small, and probably unobservable.
The circular 
polarization is predicted to be larger for so-called infrared dark 
clouds (IRDCs), although it is still small. 
For protoplanetary disks the circular polarization may be measurable, 
but will depend on how the
direction of grain alignment changes in the disk.

The paper is organized as follows.  The equations describing propagation of
partially-polarized radiation are presented in Section 
\ref{sec:radiative transfer}, and
the optics of partially-aligned dust mixtures
are summarized in Section \ref{sec:dust}.
Section \ref{sec:clouds} estimates the circularly polarized emission from
molecular clouds, including IRDCs.
Section \ref{sec:disks} discusses the alignment of solid particles in
stratified protoplanetary disks resembling HL Tau.
If the grain alignment is due to dust-gas streaming, the emission may
be circularly-polarized.
The results are discussed in Section \ref{sec:discussion}, 
and summarized in Section \ref{sec:summary}.

\section{\label{sec:radiative transfer} 
         Polarized Radiative Transfer}

\subsection{Refractive Index of a Dusty Medium}

Aligned dust grains result in linear dichroism -- the attenuation
coefficient depends on the linear polarization of the radiation.
Linear dichroism is
responsible for the polarization of starlight -- initially unpolarized
light from a star becomes linearly polarized as the result of
polarization-dependent attenuation by aligned dust grains.

We adopt the convention that the electric field
$E \propto {\rm Re}[e^{imkz-i\omega t}]$ for a wave
propagating in the $+\bzhat$ direction, where
$k\equiv\omega/c = 2\pi/\lambda$ is the wave vector
{\it in vacuo}, and $m(\omega)$ 
is the complex refractive index of the dusty medium.
For radiation polarized with $\bE\parallel\behat_j$,
the complex refractive index is
\beq
m_j \equiv 1 + m_j^\prime + i m_j^{\prime\prime}
~~~.
\eeq
The real part $m_j^\prime$ describes retardation of the wave,
relative to propagation {\it in vacuo}.  The phase delay $\phi$ varies as
\beq
\frac{d\phi_j}{dz} =
\frac{2\pi}{\lambda} m_j^\prime = n_d C_{{\rm pha},j}
~~~,
\eeq
where $n_d$ is the number density of dust grains, and
$C_{{\rm pha},j}$ is the ``phase shift'' cross section of a grain.
The imaginary part $m_j^{\prime\prime}$ describes attenuation of the energy
flux $F$:
\beq
\frac{d\ln F}{dz} = 
- \frac{4\pi}{\lambda} m_j^{\prime\prime} = -n_d C_{{\rm ext},j}
~~~,
\eeq
where $C_{{\rm ext},j}$ is the extinction cross section.

\subsection{Transfer Equations for the Stokes Parameters}

Consider a beam of radiation characterized by the usual
Stokes vector ${\bf S}\equiv(I,Q,U,V)$.  
The equations describing transfer of radiation through a dichroic 
and birefringent medium with
changing magnetic field direction have
been discussed by \citet{Serkowski_1962} and
\citet{Martin_1974}.\footnote{
   Our axes $\bxhat$ and $\byhat$ correspond, respectively, 
   to axes 2 and 1 in \citet{Martin_1974}.}
The discussions have asssumed that the aligned grains 
polarize the light by preferential attenuation of one of the polarization
modes, with circular polarization then arising from differences in propagation
speed of the linearly polarized modes.

For submicron particles, scattering is negligible
at far-infrared wavelengths, because the grain is small compared to the
wavelength.  
However, the grains are themselves able to radiate, 
and aligned grains will emit polarized radiation.

Let the direction of the static magnetic field $\bB_0$ be
\beq
\bbhat \equiv \frac{\bB_0}{|\bB_0|} = 
(\hat{\bf n}\cos\Psi  + \hat{\bf e}\sin\Psi )\sin\gamma
+ \bzhat \cos\gamma 
\eeq
where $\hat{\bf n}$ and $\hat{\bf e}$ are unit vectors in the North and East 
directions, $\bzhat =  \hat{\bf n}\times\hat{\bf e}$
is the direction of propagation, and $\sin\gamma=1$ if $\bbhat$ is in the
plane of the sky.

\begin{figure}
\begin{center}
\includegraphics[angle=0,width=6.0cm,
                 clip=true,trim=0.5cm 5.0cm 0.5cm 2.5cm]
{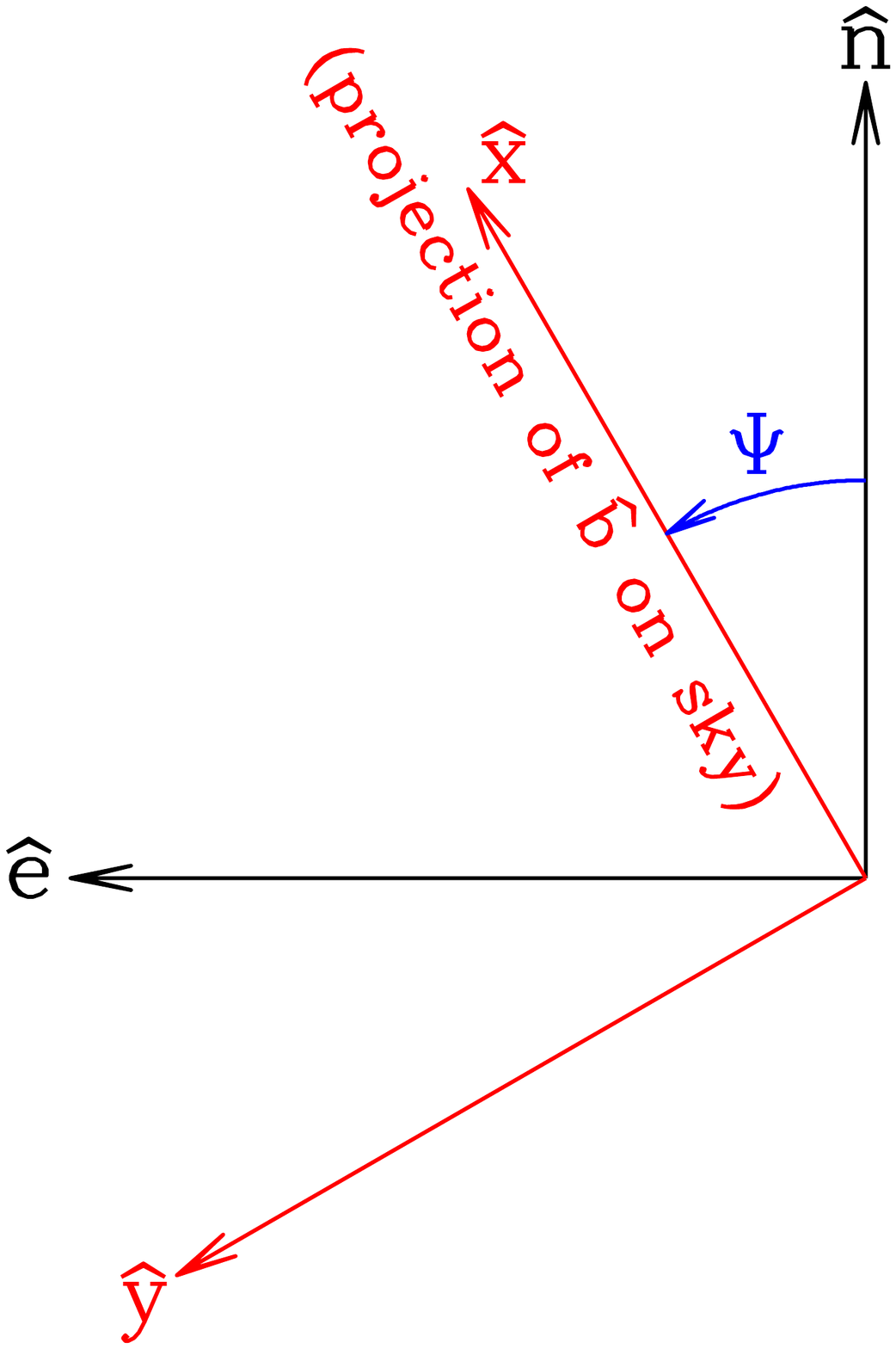}
\caption{\footnotesize\label{fig:coords}
    Angle $\Psi$, and directions $\bxhat$, $\byhat$.
    \btdnote{f1.pdf}}
\end{center}
\end{figure}

Let $\bxhat$ and $\byhat$ be orthonormal vectors in the plane of the sky, with
$\bxhat$ parallel to the projection of $\bB_0$ on the plane of the sky
(see Figure \ref{fig:coords}):
\beqa
\bxhat &\,=\,& \hat{\bf n} \cos\Psi  + \hat{\bf e} \sin\Psi 
\\
\byhat &=& -\hat{\bf n} \sin\Psi + \hat{\bf e} \cos\Psi 
~.
\eeqa

If the dust grains are partially 
aligned with their short axes tending to be parallel to $\bB_0$, we expect
$C_{{\rm ext},y} > C_{{\rm ext},x}$.
At long wavelengths ($\lambda \gg 10\micron$) we also expect
$C_{{\rm pha},y} > C_{{\rm pha},x}$.
We assume that the dust grains themselves have no overall chirality, hence
circular dichroism and circular birefringence can be neglected so long
as the response of the magnetized plasma is negligible, which is
generally the case for $\nu \gtsim 30\GHz$.

Following the notation of \citet{Martin_1974}, define
\beqa
\delta &\,\equiv\,& 
n_d ~
\frac{(C_{{\rm ext},y}+C_{{\rm ext},x})}{2} 
= \frac{2\pi}{\lambda} \left(m_x^{\prime\prime} + m_y^{\prime\prime}\right)
\\
\Delta\sigma 
&\equiv& 
n_d ~
\frac{(C_{{\rm ext},y}-C_{{\rm ext},x})}{2}
= 
\frac{2\pi}{\lambda}\left(m_y^{\prime\prime} - m_x^{\prime\prime}\right)
\\
\Delta\epsilon 
&\equiv& 
n_d ~ 
\frac{(C_{{\rm pha},y}-C_{{\rm pha},x})}{2}
= 
\frac{2\pi}{\lambda}\frac{\left(m_y^\prime - m_x^\prime\right)}{2}
~.
\eeqa
If scattering is neglected, 
the propagation of the Stokes parameters is given 
by\footnote{Eq.\ (\ref{eq:propagation}) conforms to the IEEE and 
IAU conventions for the Stokes parameters
\citep{Hamaker+Bregman_1996}: $Q>0$ for $\bE$ along the N-S direction,
$U>0$ for $\bE$ along the NE-SW direction, $V>0$ for ``right-handed''
circular polarization
($\bE$ rotating in the counterclockwise direction as viewed on the sky).}
\beq \label{eq:propagation}
\frac{d}{dz}
\left(
\begin{array}{c}
I \\ Q \\ U \\ V \\
\end{array}
\right)
=
\left(
\begin{array}{c c c c}
-\delta               & \Delta\sigma\cos2\Psi & \Delta\sigma\sin2\Psi & 0 \\
\Delta\sigma\cos2\Psi & -\delta               & 0                     & \Delta\epsilon\sin2\Psi \\
\Delta\sigma\sin2\Psi & 0                     & -\delta               & -\Delta\epsilon\cos2\Psi \\
0                     & -\Delta\epsilon\sin2\Psi & \Delta\epsilon\cos2\Psi & -\delta\\
\end{array}
\right)
\left(
\begin{array}{c}
I-B(T_d)\\ Q \\ U \\ V \\
\end{array}
\right)
~,
\eeq
where 
$B(T_d)$ is the intensity of blackbody radiation
for dust temperature $T_d$.
Eq.\ (\ref{eq:propagation}) 
differs from \citet{Martin_1974} only by 
replacement of $I$ by $(I-B)$ on the
right-hand side to allow for thermal emission
\citep[see also][]{Reissl+Wolf+Brauer_2016}. 
It is apparent that Eq.\ (\ref{eq:propagation})
is consistent with thermal equilibrium blackbody radiation, with $d{\bf S}/dz=0$
for
${\bf S}=(B,0,0,0)$.

\section{\label{sec:dust}
         Optical Properties of the Dust}

We now assume that the grains can be
approximated by spheroids.
\citet{Draine+Hensley_2021c} found that observations of
starlight polarization and
far-infrared polarization appear to be
consistent with dust with oblate spheroidal shapes, with
axial ratio $b/a\approx 1.6$ providing a good fit to observations.
\added{Observations of the diffuse ISM are consistent with
\beq
\label{eq:opacity}
\frac{\delta}{\nH} = \frac{\tau}{\NH} \approx 6.5\times10^{-27} 
\left(\frac{\lambda}{\mm}\right)^{-1.8} \cm^2\Ha^{-1}
\eeq
for $100\micron \ltsim \lambda \ltsim 1\cm$ 
\citep{Hensley+Draine_2021a,Draine+Hensley_2021a}.}

Let $\bbhat$ be a ``special'' direction in space for grain alignment: the short
axis $\bahat_1$ of the grain
may be preferentially aligned either parallel or perpendicular
to $\bbhat$.  For grains in the diffuse ISM, $\bbhat$ is the
magnetic field direction, and the short axis $\bahat_1$ tends to be
parallel to $\bbhat$.  In protostellar disks, however, other alignment
mechanisms may operate, and $\bbhat$ may not be parallel to the magnetic
field.

We \replaced{will assume}{approximate} 
the grains \replaced{to be}{by} oblate spheroids, spinning with short axis
$\bahat_1$ parallel to the angular momentum $\bJ$.
For oblate spheroids,
the fractional alignment is defined to be 
\beq \label{eq:falign}
\falign\equiv 
\frac{3}{2}
\langle (\bahat_1\cdot\bbhat)^2\rangle -
\frac{1}{2}
~~~,
\eeq
where $\langle ...\rangle$ denotes averaging over the grain population.
If $\bJ\parallel\bbhat$, then $\falign\rightarrow1$; 
if $\bJ$ is randomly-oriented, then $\falign=0$;
if $\bJ\perp\bbhat$, then $\falign\rightarrow-\frac{1}{2}$.

The ``modified picket fence approximation'' \citep{Draine+Hensley_2021c}
relates $\delta$, $\Delta\sigma$,
and $\Delta\epsilon$ to $\falign$ and the angle $\gamma$: 
\beqa \label{eq:delta from MPFA}
\delta &~=~& n_d 
\left[ \frac{C_{{\rm abs},a}+2C_{{\rm abs},b}}{3}
        + \falign\left(\cos^2\gamma-\frac{1}{3}\right)
          \frac{\left(C_{{\rm abs},b}-C_{{\rm abs},a}\right)}{2}
\right]
\\
\Delta \sigma &=& n_d \falign\sin^2\gamma ~
       \frac{\left(C_{{\rm abs},b} - C_{{\rm abs},a}\right)}{2}
\\
\Delta\epsilon &=& n_d \falign\sin^2\gamma ~
       \frac{\left(C_{{\rm pha},b} - C_{{\rm pha},a}\right)}{2}
~~~.
\eeqa
In the Rayleigh limit (grain radius $a\ll\lambda$) we have
\citep{Draine+Lee_1984}
\beqa
C_{{\rm abs},j} &~=~& 
\frac{2\pi V}{\lambda}
\frac{\epsilon_2}
     {|1+(\epsilon-1)L_j|^2}
\\
C_{{\rm pha},j} &~=~& 
\frac{\pi V}{\lambda}
\frac{\left\{
     (\epsilon_1-1)\left[1+L_j(\epsilon_1-1)\right]+\epsilon_2^2L_j
     \right\}
     }
     {|1+(\epsilon-1)L_j|^2}
~~~,
\eeqa
where $\epsilon(\lambda)\equiv\epsilon_1+i\epsilon_2$ is the complex dielectric
function of the grain material, and
$L_a$ and $L_b=(1-L_a)/2$ are dimensionless ``shape factors''
\citep{van_de_Hulst_1957,Bohren+Huffman_1983} that depend on the
axial ratio of the spheroid.
\citet{Draine+Hensley_2021a} have estimated 
$\epsilon(\lambda)$ of astrodust for 
different assumed axial ratios.

\begin{figure}
\begin{center}
\includegraphics[angle=0,width=10.0cm,
                 clip=true,trim=0.5cm 5.0cm 0.5cm 2.5cm]
{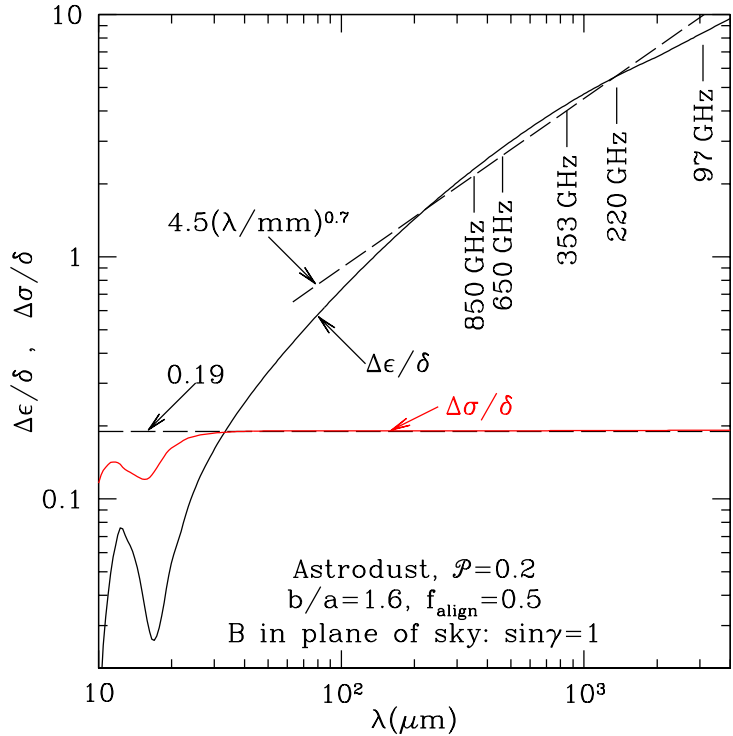}
\caption{\label{fig:Cpha/Cabs} 
The ratios $\Delta\sigma/\delta$ and $\Delta \epsilon/\delta$
for oblate astrodust
spheroids with porosity $\calP=0.2$, axial ratio $b/a=1.6$,  
alignment fraction 
$\falign=0.5$, and $\sin^2\gamma=1$ (magnetic field in the plane of the
sky).
The power-law approximation (\ref{eq:powerlaw}) for $\Delta\epsilon/\delta$
is also shown. 
\btdnote{f2.pdf}
}
\end{center}
\end{figure}

Figure \ref{fig:Cpha/Cabs} shows the 
dimensionless ratios $\Delta\sigma/\delta$ and $\Delta\epsilon/\delta$
for oblate astrodust spheroids with \added{porosity $\calP=0.2$,} $b/a=1.6$ 
($L_a=0.464$, $L_b=0.268$)
and
$\falign=0.5$, for the case where the magnetic field is in the plane of
the sky ($\sin\gamma=1$).
The relatively high opacity that enables ``astrodust'' to reproduce the
observed far-infrared emission and polarization also implies that $\epsilon_1$
has to be fairly large at long wavelengths
\citep{Draine+Hensley_2021a}.
This causes $\Delta\epsilon/\delta$ to be relatively large, as seen in 
Figure \ref{fig:Cpha/Cabs}.
For $\lambda \gtsim 70\micron$, oblate astrodust grains with $b/a=1.6$ have
\beqa \label{eq:Delta sigma/delta in FIR}
\frac{\Delta\sigma}{\delta} 
&\,\approx\,& 
0.38 \falign \sin^2\gamma
\\ \label{eq:powerlaw}
\frac{\Delta\epsilon}{\delta}
&\approx&
9.0\left(\frac{\lambda}{\mm}\right)^{0.7}
\falign \sin^2\gamma
~~~.
\eeqa
Eq.\ (\ref{eq:Delta sigma/delta in FIR}) and (\ref{eq:powerlaw})
neglect the weak dependence of $\delta$ on $\falign$ and $\gamma$
(see Eq.\ \ref{eq:delta from MPFA}).
Eqs.\ (\ref{eq:Delta sigma/delta in FIR}) and
((\ref{eq:powerlaw}) are shown in Figure \ref{fig:Cpha/Cabs} for
$\falign\sin^2\gamma=0.5$.

\section{\label{sec:clouds}
         Circular Polarization from Interstellar Clouds}

\subsection{Grain Alignment}

A spinning grain develops a magnetic moment from the Barnett effect
(if it has unpaired electrons) 
and the Rowland effect (if it has a net charge).
For submicron grains, the resulting net magnetic
moment is large enough that
the Larmor precession period in the local interstellar magnetic field is
short compared to the timescales for other mechanisms 
to change the direction of
the grain's angular momentum $\bJ$.  The rapid 
precession of $\bJ$ around 
the local magnetic field $\bB_0$ and the resulting averaging of grain
optical properties 
establishes $\bB_0$ as the special direction for grain alignment -- 
grains will
be aligned with their short axis preferentially oriented either parallel or
perpendicular to $\bB_0$.  

Paramagnetic dissipation,
radiative torques, or systematic streaming of the grains relative to the
gas will determine whether the grains 
align with their short axes preferentially
parallel or perpendicular to $\bB_0$.  Although the details of the physics of 
grain alignment are not yet fully understood, it is now
clear
that grains in diffuse and translucent clouds tend to align with short axes
$\bahat_1$ tending to be parallel to $\bB_0$, i.e., with
$\falign>0$ (see Eq.\ \ref{eq:falign}).

If the dust
grains are modeled by oblate spheroids with axial ratio $b/a=1.6$, 
a mass-weighted alignment fraction $\falign\approx 0.5$ 
can reproduce the
highest observed levels of polarization of both starlight and far-infrared
emission from dust in diffuse clouds (including
diffuse molecular clouds) \citep{Draine+Hensley_2021c}.

In dark clouds, the fractional polarization of the thermal emission is
generally lower than in diffuse clouds.  
The lower fractional polarization may indicate
lower values of $\falign$ within dark clouds,
but it could also 
result from a nonuniform
magnetic field in the cloud, with the overall linear polarization fraction
reduced by beam-averaging over regions with different
polarization directions.

If the reduced values of linear polarization are 
due to \added{systematic} 
changes in magnetic field direction along the line-of-sight, 
the emission from the cloud could 
become partly circularly-polarized.  We now estimate
what levels of circular polarization might be present.
 
\subsection{\label{sec:nearby MCs}
            Nearby Molecular Clouds}

Planck has observed linearly polarized emission from many molecular
clouds.
To estimate the levels of circular polarization that might be present,
we consider one 
illustrative example, in the ``RCrA-Tail'' region in the R Corona Australis
molecular cloud
\citep[see Fig.\ 11 in][]{Planck_int_results_xix_2015}.
The polarized emission in this region has a number of local maxima.
One of the polarized flux maxima coincides with a total emission peak
near
$(\ell,b)\approx (-0.9^\circ,-18.7^\circ)$,
with total intensity $I(353\GHz)\approx 4\MJy\sr^{-1}$ and 
linear polarization fraction $p\approx 2.5\%$.

For an assumed dust temperature
$T_d\approx 15\K$, the observed intensity
$I(353\GHz)=4\MJy\sr^{-1}$ implies 
$\tau(353\GHz)\approx 1.3\times10^{-4}$.
For diffuse ISM dust \citep[see, e.g.][]{Hensley+Draine_2021a}, this would
correspond to $A_V\approx 5\,$mag.


For simple assumptions about the angle $\Psi$ characterizing the
projection of the magnetic field on the sky, we can obtain approximate
analytic solutions to the radiative transfer equations
(\ref{eq:propagation}), valid for $\tau \ll 1$
(see Appendix \ref{app:uniform twist}).  
Define $d\tau^\prime \equiv \delta dz$.
Suppose that $T_d$, $(\Delta\sigma/\delta)$,
and $(\Delta\epsilon/\delta)$ are constant, and
assume that the
magnetic field direction has a smooth twist along the line of sight,
with $\Psi$ varying linearly with $\tau^\prime$ as $\tau^\prime$ varies
from $0$ to $\tau$:
\beq \label{eq:linear Psi}
\Psi = \Psi_0 + \alpha\tau^\prime ~~~,~~~~\alpha\equiv \frac{\Delta\Psi}{\tau}
~~~.
\eeq

For $\tau\ll 1$, 
the linear and circular polarization fractions are then
(see Appendix \ref{app:uniform twist})
\beqa \label{eq:p approx}
p &~\approx~&
\left(\frac{\Delta\sigma}{\delta}\right)
\frac{\left[1-\cos(2\Delta\Psi)\right]^{1/2}}{\Delta\Psi}
\\ \label{eq:V/I approx}
\frac{V}{I} &\approx& 
\left(\frac{\Delta\sigma}{\delta}\right)
\left(\frac{\Delta\epsilon}{\delta}\right)
\frac{\tau}{2\Delta\Psi}
\left[1 - \frac{\sin(2\Delta\Psi)}{2\Delta\Psi}\right]
~~~.
\eeqa
Eq.\ (\ref{eq:p approx}--\ref{eq:V/I approx}) are for the special case of
an isothermal medium with a uniform twist in the alignment direction.

If we assume diffuse cloud dust properties (Eq.\ \ref{eq:Delta sigma/delta in FIR}, \ref{eq:powerlaw}) but with
$\falign\sin^2\gamma=0.075$
and a twist angle $\Delta\Psi=90^\circ$,
%
%
%
we can reproduce the observed polarization $p\approx 2.5\%$
in the RCrA-Tail region.
With these parameters, Eq. (\ref{eq:V/I approx}) predicts
circular polarization
\replaced{$V/I \approx 7\times10^{-7}$}
{$V/I \approx 7\times10^{-7}
(\lambda/850\micron)^{-1.1}$}, 
far below
current sensitivity limits.
It is clear that
measurable levels of circular polarization in the far-infrared
will require much larger optical 
depths $\tau$.


\subsection{Infrared Dark Clouds}

\begin{figure}
\begin{center}
\includegraphics[angle=0,width=8.0cm,
                 clip=true,trim=0.5cm 5.0cm 0.5cm 2.5cm]
{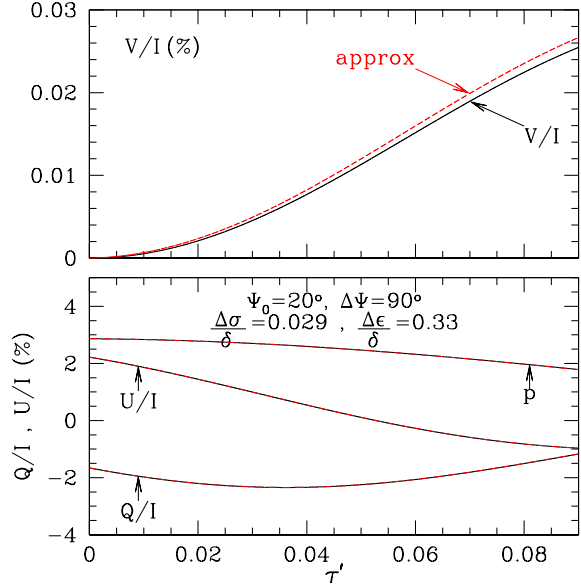}
\caption{\label{fig:stokes params}\footnotesize
Stokes parameters for radiation from a dust slab resembling
the Brick IRDC, as a function of
the optical depth $\tau^\prime$
along the path.
The properties of astrodust at $850\GHz$ ($350\micron$)
are assumed, with 
\replaced{$\falign\sin^2\gamma=0.12$}{$\falign\sin^2\gamma=0.075$},
and a field rotation $\Delta\Psi=90^\circ$.
\added{Black curves: numerical results.  Red curves: analytic approximations
(\ref{eq:I(tau)}--\ref{eq:p}).}
\btdnote{f3.pdf}
}
\end{center}
\end{figure}

Typical giant molecular clouds \added{(GMCs)}, 
such as the Orion Molecular Cloud,
have mass surface densities resulting in
$A_V \approx 10{\rm\,mag}$ of extinction, and are therefore
referred to as ``dark clouds''.
However, in the inner Galaxy, a number of clouds have
been observed that appear to be ``dark'' (i.e., opaque) \added{even} in the
mid-infrared.
These ``infrared dark clouds'' (IRDCs) have
dust masses per area an order of magnitude larger
than ``typical'' giant molecular clouds.
Because of the much larger extinction in IRDCs, the circular
polarization may be much larger than in normal GMCs.

The ``Brick'' (G0.253+0.016) is a well-studied IRDC
\citep{Carey+Clark+Egan+etal_1998,
Longmore+Rathborne+Bastian+etal_2012}.
With an estimated mass $M>10^5\Msol$ and high estimated density
($\nH>10^4\cm^{-3}$), the Brick appears to be forming stars
\citep{Marsh+Ragan+Whitworth+Clark_2016,
       Walker+Longmore+Bally+etal_2021}, although with
no signs of high-mass star formation.
It has been mapped at $70-500\micron$ by Herschel Space Observatory
\citep{Molinari+Schisano+Elia+etal_2016} and at 
$220\GHz$ by ACT
\citep{Guan+Clark+Hensley+etal_2021}.
Polarimetric maps have been made at $220\GHz$ by ACT, and at $850\GHz$
by the CSO \citep{Dotson+Vaillancourt+Kirby+etal_2010}.

The Northeastern region at $(\ell,b)=(16^\prime,2^\prime)$
has $I(600\GHz)\approx 5000\MJy\sr^{-1}$
\citep{Molinari+Schisano+Elia+etal_2016}
and $I(220\GHz)\approx 90\MJy\sr^{-1}$
\citep{Guan+Clark+Hensley+etal_2021}. 
For an assumed
dust temperature $T_d\approx 20\K$, this
indicates optical depths
$\tau(600\GHz)\approx 0.05$, $\tau(220\GHz)\approx0.005$.
Astrodust would then have 
$\tau(850\GHz)\approx 0.09$, and $\tau(353\GHz) \approx 0.014$ -- 
about 100 times
larger than in the R CrA molecular cloud.

The fractional polarization is expected to be approximately independent
of frequency in the submm.
At $220\GHz$,
\citep{Guan+Clark+Hensley+etal_2021} report 
a linear polarization 
of \replaced{$3.6\%$}{$1.8\%$} 
at 220 GHz for the Northeastern end of the cloud, 
\added{$(\ell,b)\approx(16^\prime,2.5^\prime)$ 
(Yilun Guan 2021, private communication).}
The CSO polarimetry suggests a \replaced{smaller}{similar} 
fractional polarization \added{at 850\,GHz}.
\deleted{We take the fractional polarization at
$(\ell,b)=(16^\prime,2^\prime)$
to be $\sim$3\%.}

While this fractional polarization is relatively small compared to the
highest values ($\sim 20\%$) observed by Planck in diffuse clouds, it is
still appreciable, requiring significant grain alignment in a substantial
fraction of the cloud volume 
(i.e., not just in the surface layers of the IRDC).
The inferred average magnetic field direction $\Psi\approx 20^\circ$
\citep{Guan+Clark+Hensley+etal_2021}
differs by
$\sim$$60^\circ$ from the $\Psi\approx80^\circ$ field direction
indicated by the $220\GHz$ polarization
outside the cloud, demonstrating that the magnetic field in this region
is far from uniform.

As a simple example,
we suppose, as we did for the RCrA-Tail region above, that
the projected field rotates 
by $\Delta\Psi=90^\circ$ from the far side of the ``Brick'' to the near side.

We calculate the 
circular polarization
at $850\GHz$ ($350\micron$)
for the estimated total optical depth $\tau(850\GHz)=0.09$ of the Brick.
We use the estimated properties of astrodust in the diffuse ISM, 
with $\falign\sin^2\gamma=$\replaced{$0.12$}{$0.075$} 
to approximately reproduce the $\sim$\replaced{$3\%$}{$1.8\%$} 
polarization observed for the Brick.

Figure 2 shows the polarization state of the radiation as it propagates
through the cloud from $\tau^\prime=0$ to $\tau^\prime=\tau$.  The fractional
polarization $p$ starts off at $\sim$\replaced{4.7}{2.9}\%, 
dropping to $\sim$\replaced{3\%}{1.8\%} at
$\tau^\prime=\tau$ as the result of the assumed magnetic field twist of
$\Delta\Psi=90^\circ$.

The resulting $850\GHz$ 
circular polarization $V/I$ is small, only 
$\sim$\replaced{$0.065\%$}{$0.025\%$}.
Measuring such low levels of circular polarization will be challenging.
For $\Delta\epsilon/\delta \propto \lambda^{0.7}$ 
(see Figure \ref{fig:Cpha/Cabs}) and the absorption coefficient
$\delta \propto \lambda^{-1.8}$
\added{(see Eq.\ \ref{eq:opacity})}, the circular polarization from an IRDC
is expected to vary as $V/I \propto \lambda^{-1.1}$.
For the adopted parameters ($\Delta\Psi=90^\circ$, 
$\tau(850\GHz)=0.09$, 
$\falign\sin^2\gamma=$\replaced{$0.12$}{$0.075$}), the Brick would have
\beq
\frac{V}{I} \approx \replaced{0.065}{0.025}\%\left(\frac{350\micron}{\lambda}\right)^{1.1}
~~~
\eeq
\added{for $70\micron \ltsim \lambda \ltsim 1\cm$.}
While much larger than for normal GMCs, this estimate for the
circularly-polarized emission from the Brick is \deleted{still} small, and
measuring it will be challenging.  

\added{\subsection{PDR Seen Through a Molecular Cloud}}

\begin{figure}
\begin{center}
\includegraphics[angle=0,width=10.0cm,
                 clip=true, trim=0.5cm 0.5cm 0.5cm 0.5cm]
{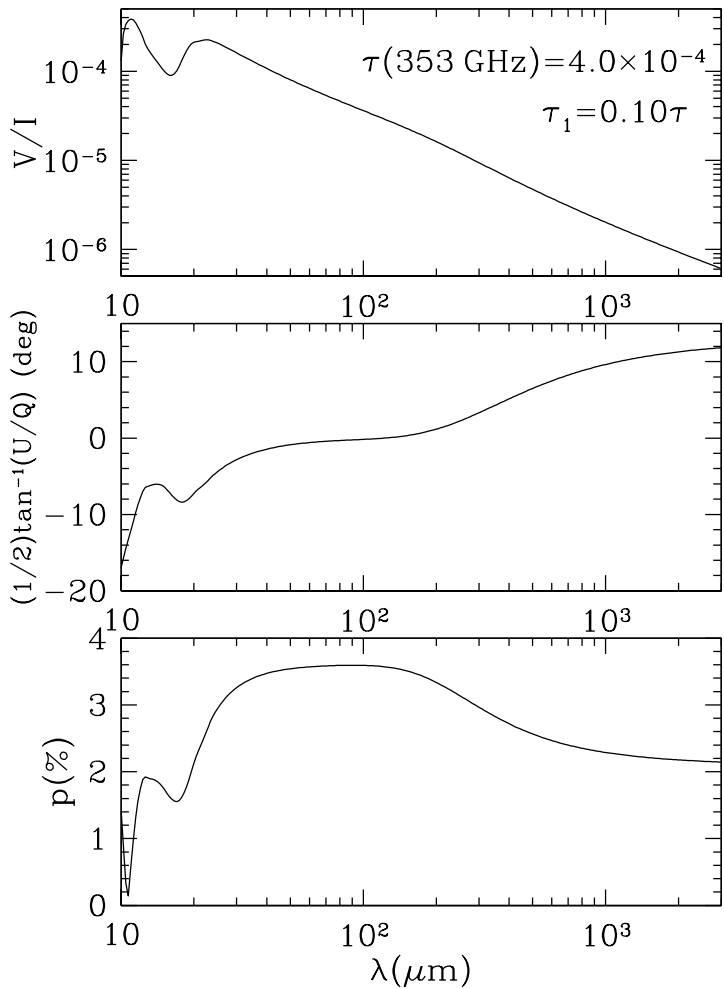}
\caption{\footnotesize\label{fig:pdrspec}
    \added{
    Polarization of dust continuum from a molecular cloud with a
    PDR on the far side.
    The magnetic field in the cold cloud is assumed to have a systematic
    twist along the line of sight, with a twist angle
    $\Delta\Psi=60^\circ$.
    The dust is assumed to be partially aligned, with
    $\falign\sin^2\gamma=0.1$ for the warm dust ($T_1=80\K$) in the PDR,
    and $\falign\sin^2\gamma=0.05$ for the cold dust ($T_2=15\K$) 
    in the rest of the cloud.
   }}
\end{center}
\end{figure}

\added{The warm dust surrounding 
an embedded HII region may allow measurement of circular polarization
at wavelengths as short as $\sim$$20\micron$.
We consider a cloud with optical depth $\tau(353\GHz)=4\times10^{-4}$,
somewhat greater than the R Corona Australis cloud example considered in Section
\ref{sec:nearby MCs}, but small compared to the Brick.

The far edge of the cloud is assumed to contain 
warm dust in a photodissociation region (PDR)
with optical depth
$\tau_1(\lambda)$.
The PDR is assumed to contribute
10\% of the total column density through the
the molecular cloud, with
dust heated to $T_1=80\K$.
The dust
in the rest of the molecular cloud is cold, $T_2=15\K$.

We assume the dust in the PDR to be moderately aligned, with
$\falign\sin^2\gamma=0.1$, whereas for the dust in the rest of the molecular
cloud we take $\falign\sin^2\gamma=0.05$.

Using the analytic approximation for the ``two-zone'' model 
in Appendix \ref{app:two zone}, we find the 
the fractional linear polarization $p$ and circular polarization $V/I$ shown in 
Figure \ref{fig:pdrspec}.  
At $\lambda \ltsim 100\micron$, the
polarization is the combination of polarized emission
from the warm dust in the PDR and dichroic absorption by the cool dust.
At longer wavelengths, $\lambda > 300\micron$, dichroic absorption is minimal,
and we see the sum of the polarized emission from the warm and cool regions.
The polarization angle rotates  as the ratio of warm emission
to cool emission drops with increasing wavelength.
The features at $10< \lambda < 30\micron$ arise from the strong silicate
absorption bands at $10$ and $18\micron$.

The circular polarization reaches 
$V/I=0.02\%$ at $\lambda=20\micron$
but declines as $\sim\!\lambda^{-1.1}$ at longer wavelengths.
}

\bigskip

\section{\label{sec:disks}
         Circular Polarization from Protoplanetary Disks}

Protoplanetary disks can have dust surface densities well in excess of
IRDCs, raising the possibility that $\tau$ may be large enough to generate
measurable circular polarization if the grains are locally aligned 
{\it and} the
alignment direction varies along the optical path.

\subsection{Grain Alignment in Protoplanetary Disks}

Gas densities in protoplanetary disks exceed interstellar gas densities by
many orders of magnitude.
The observed thermal emission spectra from young protoplanetary
disks appear to require that
most of the solid material be in particles
with sizes that may be as large as $\sim$mm
\citep{Beckwith+Sargent_1991,Natta+Testi_2004,Draine_2006a},
orders of magnitude larger than \added{the submicron grains} 
in the diffuse ISM.

The physics of grain alignment in protoplanetary
disks differs substantially
from the processes in the diffuse ISM.
One important difference from interstellar clouds
is that in protoplanetary disks the Larmor precession period 
for the grain sizes of interest is {\it long} 
compared to the time for the grain
to undergo collisions with a mass of gas atoms equal to the grain mass
\citep{Yang_2021}.
With Larmor precession no longer important, 
the magnetic field no longer determines the preferred direction
for grain alignment.  Instead, the ``special'' direction
may be either the local direction of
gas-grain streaming -- in which case, $\bbhat\parallel\bvdrift$ -- or
perhaps the direction of anisotropy in the radiation field
-- in which case, $\bbhat\parallel\br$.  Whether grains will tend to align
with short axes $\bahat_1$ parallel or perpendicular to $\bbhat$ 
\added{(i.e., $\falign>0$ or $\falign<0$)} is a separate
question.

\subsubsection{Alignment by Radiative Torques?}

Radiative torques resulting from outward-directed radiation provide 
\replaced{another}{one}
possible mechanism for grain alignment.
Starlight torques have been found to be very important for both
spinup and alignment of interstellar grains
\citep{Draine+Weingartner_1996,
       Draine+Weingartner_1997,
       Weingartner+Draine_2003,
       Lazarian+Hoang_2007a}.
With mm-sized grains, both stellar radiation and infrared emission from the
disk may be capable of exerting systematic torques large enough to
affect the spin of the grain.  However, the radiation pressure
$\sim L_\star/4\pi R^2c \approx 5\times10^{-9}(L_\star/\Lsol)(100\AU/R)^2\erg\cm^{-3}$ is small compared to the gas pressure
$\sim 8\times10^{-5}(\nH/10^{10}\cm^{-3})(T/100\K)\erg\cm^{-3}$.
If the grain streaming velocity exceeds $\sim 10^{-4} c_s$, where $c_s$
is the sound speed, systematic torques exerted by gas atoms may dominate
radiative torques.
Studies of realistic grain geometries are needed to clarify the relative
importance of gaseous and radiative torques.

\subsubsection{Alignment by Grain Drift?}

The differential motion of dust and gas in three-dimensional disks
has been discussed by
\citet{Takeuchi+Lin_2002}.
Grains well above or below
the midplane will 
sediment toward the midplane, with
$\bvdrift \parallel \bz_{\rm disk}$, where
$z_{\rm disk}$ is height above the midplane.
Dust grains close to the midplane will be in near-Keplerian orbits, but will
experience a ``headwind'', with $\bvdrift \parallel \hat{\phi}$.
Vertical and azimuthal drift velocities will in general differ, with different
dependences on grain size and radial distance from the protostar.

\citet{Gold_1952} proposed  
grain drift relative to the gas as an alignment mechanism.
For hypersonic motion,
Gold concluded that needle-shaped particles would tend to align
with their short axes perpendicular to 
$\bvdrift$.
\citet{Purcell_1969a} analyzed spheroidal shapes, finding that significant
alignment requires hypersonic
gas-grain velocities 
if the grains are treated as rigid bodies.
The degree of grain alignment \added{of spheroidal grains} 
is increased when dissipative processes
within the grain are included \citep{Lazarian_1994b}, but the degree of
alignment is small unless the streaming is supersonic.

\citet{Lazarian+Hoang_2007b} discussed mechanical alignment of
subsonically-drifting grains with ``helicity'', arguing that helical
grains would preferentially acquire angular momentum parallel or antiparallel
to $\bvdrift$; internal dissipation would then cause the short
axis to tend to be {\it parallel} to $\bvdrift$.
\citet{Lazarian+Hoang_2007b} based their analysis on a simple geometric 
model of a spheroidal grain with a single projecting panel.
More realistic irregular geometries have been considered by
\citet{Das+Weingartner_2016} and \citet{Hoang+Cho+Lazarian_2018}.
However, these studies all assumed Larmor
precession to be rapid compared to the gas-drag time, and are therefore
not directly
applicable to protoplanetary disks.

It appears possible that, averaged over
the ensemble of irregular grain shapes, 
the net effect of gas-grain streaming in protoplanetary disks may
be (1) suprathermal angular momenta tending to be perpendicular to
$\bvdrift$, and (2) tendency of grains to align with short axes
perpendicular to $\bvdrift$.
Below, we consider the
consequences of this conjecture.

\subsection{The HL Tau Disk as an Example}

ALMA has observed a number of protoplanetary disks 
\citep[e.g.,][]{Andrews+Huang+Perez+etal_2018}.  HL Tau remains one of the
best-observed cases: it is nearby ($\sim$$140\pc$), bright, and
\deleted{only} moderately inclined ($i \approx 45^\circ$).
The optical depth in the disk is large, with 
beam-averaged $\tau(3.1\mm)\approx 0.13$ at $R\approx 100\AU$.\footnote{
   At $R\approx 100\AU$, $I_\nu(3.1\mm)\approx 1.1\times10^3\MJy\sr^{-1}$
   \citep{Kataoka+Tsukagoshi+Pohl+etal_2017,Stephens+Yang+Li+etal_2017},
   implying $\tau\approx 0.13$ if 
   the dust temperature $T_d\approx30\K$ \citep{Okuzumi+Tazaki_2019}.}
Given that the dust is visibly concentrated in rings, and the
possibility that there may be
additional unresolved substructure, the actual optical depth of the
emitting regions at $100\AU$ is likely to be larger.
 
The polarization in HL Tau has been mapped by ALMA at $870\micron$, 
$1.3\mm$, and $3.1\mm$ 
\citep{Kataoka+Tsukagoshi+Pohl+etal_2017,
       Stephens+Yang+Li+etal_2017}.
The observed polarization patterns show
considerable variation from one frequency to another, 
complicating interpretation. 
Both intrinsic polarization from
aligned grains and polarization resulting from scattering appear
to be contributing to the overall polarization.
\citet{Mori+Kataoka_2021} argue that
polarized 
emission makes a significant contribution to the polarization, 
at least at $3.1\mm$.

The $3.1\mm$ polarization pattern is generally azimuthal
\citep{Stephens+Yang+Li+etal_2017}.  If due to polarized emission,
this would require that the radiating dust
grains have
short axes preferentially oriented in the radial direction.
The alignment mechanism is unclear.

\citet{Kataoka+Okuzumi+Tazaki_2019} favor radiative torques, with the
grain's short axis assumed to be {\it parallel} to the radiative flux, in the
radial direction. 
This would be consistent with the observation that the linear
polarization tends to be in the azimuthal direction.
If radiative torques are responsible for grain alignment in protoplanetary
disks, then we do not expect the thermal emission from the disk to be
circularly polarized, because the grains in the upper and lower layers
of the disk will tend to have the same alignment direction as the grains near
the midplane. 
If there is no change in the direction of
the grain alignment along a ray, there will be no circular polarization.

Here we instead
suppose that grain alignment is dominated by gas-grain streaming due
to systematic motion of the dust grains relative to the local gas.
If we define $\bbhat \parallel \bvdrift$ we can apply the
discussion above.
As discussed above, 
we conjecture that the irregular grains align with short axes tending
to be \emph{perpendicular} to $\bvdrift$, thus $\falign<0$.

\begin{table}[t]
\footnotesize
\begin{center}
\caption{\label{tab:disk pars}A Stratified Disk Example$^{\rm a}$}
\begin{tabular}{c c c c}
         & lower layer & midplane & upper layer \\
Parameter  & $j= 1$      & $j= 2$   & $j= 3$ \\
\hline
$\tau_j$   & $0.05$        & $0.2$            & $0.05$ \\
$B(T_{d,j})/B(T_{d,2})$  & $2$           & $1$              & $2$   \\
$\sin\gamma$ & $\cos\theta_i$ & $1\rightarrow\sin\theta_i$ 
                              & $\cos\theta_i$ \\  
$\Psi_j$   & $0$           & $0\rightarrow180^\circ$ & $180^\circ$ \\
$\falign$ & $-0.2$ & $-0.2$         & $-0.2$ \\ 
$(\Delta\sigma/\delta)_j$& $0.38\falign\sin^2\gamma$  
                         & $0.38\falign\sin^2\gamma$
                         & $0.38\falign\sin^2\gamma$ \\
$(\Delta\epsilon/\delta)_j$ & $19\falign\sin^2\gamma$
                            & $19\falign\sin^2\gamma$
                            & $19\falign\sin^2\gamma$ \\
\hline 
\multicolumn{4}{l}{For $\lambda=3.1\mm$.}
\end{tabular}
\end{center}
\end{table}
\begin{figure}[b]
\begin{center}
\includegraphics[angle=0,width=7.0cm,
                 clip=true,trim=0.5cm 0.5cm 0.5cm 0.5cm]
{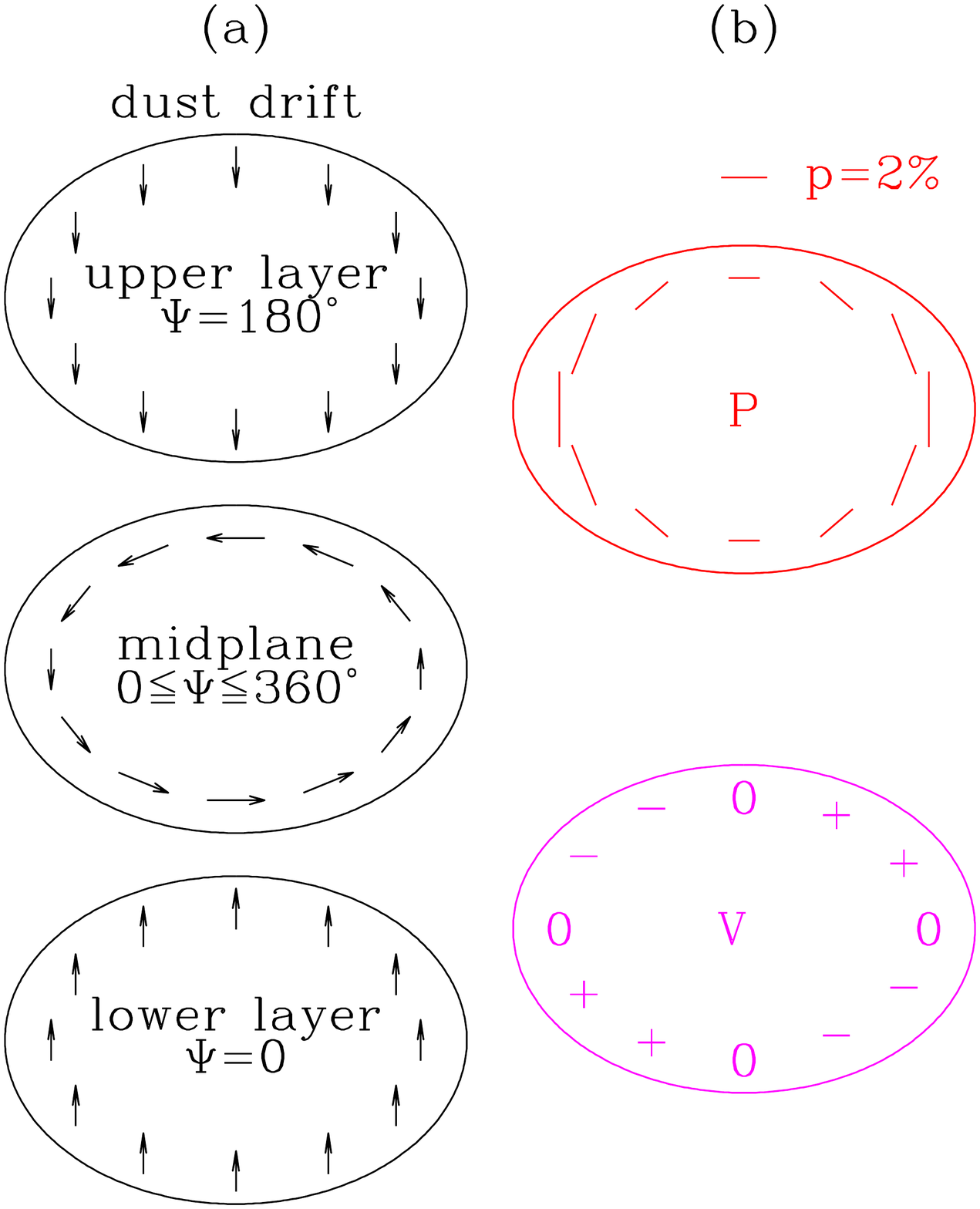}
\includegraphics[angle=0,width=7.0cm,
                 clip=true,trim=0.5cm 0.5cm 0.5cm 0.5cm]
{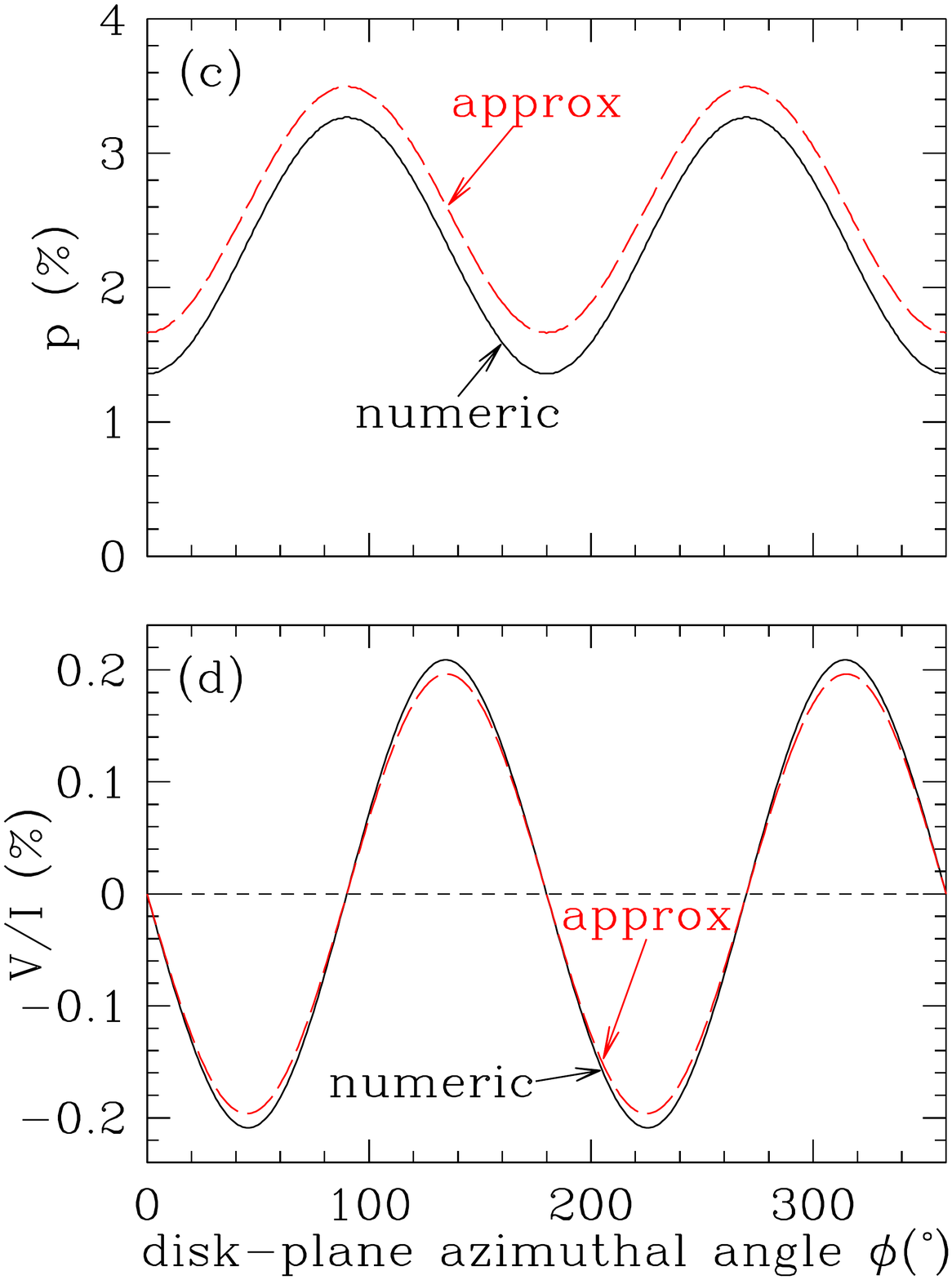}
\caption{\label{fig:disk}\footnotesize
(a) Grain drift directions in a stratified disk (see text).
(b) Upper figure: the direction of linear polarization if the
grains are aligned by $\bv_{\rm drift}$,
with parameters in Table \ref{tab:disk pars}.
The length of the line segment is proportional to the fractional
polarization, with the scale bar showing $2\%$ fractional polarization.
Lower figure:
quadrupolar pattern for circular polarization $V$ for the
example discussed in the text.
(c) Fractional linear polarization $p=(Q^2+U^2)^{1/2}/I$
for stratified disk model (see Table \ref{tab:disk pars}),
viewed at inclination $\theta_i=45^\circ$ 
(see text),
as a function of azimuthal angle in the disk plane.  $\phi=0$ is along
the minor axis.
(d) Circular polarization $V/I$ for this
model.
\btdnote{f4ab.pdf, f4cd.pdf}
}
\end{center}
\end{figure}

As before, let $\gamma$ be the angle between the line-of-sight and
$\bbhat$,
and let $\Psi$ be
the angle (relative to north) of the projection of $\bbhat$ on the
plane of the sky.
For illustration, we take the disk to have the major axis in the E-W direction
(see Figure \ref{fig:disk}), with inclination $i$.
Thus vertical drifts correspond to $\Psi=0$ and $180^\circ$.
The treatment of radiative transfer developed above for magnetized
clouds can be reapplied to protoplanetary disks --
the only difference is that if the grains align with their short
axis tending to be perpendicular to $\bvdrift$ 
then $\falign<0$, implying
$\Delta\sigma<0$ and $\Delta\epsilon<0$.

The direction and magnitude of $\bvdrift$ will vary with height in the disk.
$\bvdrift$ may be approximately normal to the disk plane
for grains that are falling toward the midplane, whereas $\bvdrift$
will be azimuthal
for grains near the midplane, with Keplerian rotation causing them to move
faster than the pressure-supported gas disk.
Thus, grain orientations may vary both vertically
and azimuthally.  With $\Psi$ varying along a ray, the emerging radiation may
be partially circularly-polarized.

The observed linear polarization of a few percent suggests that
$|\Delta\sigma/\delta| \approx $ a few \%.

We do not expect $\Psi$ to vary linearly with $\tau$ as in 
Eq.\ (\ref{eq:linear Psi}): 
the variation of $\Psi$ along the ray will depend on
the varying grain dynamics along the ray.
To investigate what levels of circular polarization might be present, 
we consider an idealized model with three dust layers:
layer 2 is the dust near the midplane,
and layers 1 and 3 contain the dust below and above the midplane.
Conditions in layers 1 and 3 are assumed to be identical.
Let $\tau_j$ be the optical depth 
through layer $j$.
Assume that $\bbhat$ is normal to the disk in layers 1 and 3, and
azimuthal in layer 2 (see Figure \ref{fig:disk}).
Thus $\Psi_1=\Psi_3$.
For small values of $\tau_1$, $\tau_2$, and $\tau_3$ we can approximate
the radiative transfer (see Appendix \ref{app:three zone model}):
\beqa
I_1 &=& B_1\tau_1 \left(1-\frac{1}{2}\tau_1\right) e^{-\tau_2-\tau_3}
\\
I_2 &=& B_2\tau_2 \left(1-\frac{1}{2}\tau_2\right) e^{-\tau_3}
\\
I_3 &=& B_3\tau_3 \left(1-\frac{1}{2}\tau_3\right)
\\
I &\,\approx\,&
I_1+I_2+I_3
\\ \label{eq:Q}
Q &\approx&
-\left(\frac{\Delta\sigma}{\delta}\right)_{\!3}
\cos(2\Psi_3)
\left[I_1+I_3-\tau_3(I_1+I_2)\right]
-\left(\frac{\Delta\sigma}{\delta}\right)_{\!2}
\cos(2\Psi_2)
\left(I_2-\tau_2I_1\right)
\\ \label{eq:U}
U &\approx&
-\left(\frac{\Delta\sigma}{\delta}\right)_{\!3}
\sin(2\Psi_3)
\left[I_1+I_3-\tau_3(I_1+I_2)\right]
-\left(\frac{\Delta\sigma}{\delta}\right)_{\!2}
\sin(2\Psi_2)
\left(I_2-\tau_2I_1\right)
\\ \label{eq:V}
V &\approx&
\sin(2\Psi_2-2\Psi_1)
\left[
\left(\frac{\Delta\epsilon}{\delta}\right)_{\!2}
\left(\frac{\Delta\sigma}{\delta}\right)_{\!1}
\tau_2 I_1
+
\left(\frac{\Delta\epsilon}{\delta}\right)_{\!3}
\left(\frac{\Delta\sigma}{\delta}\right)_{\!2}
\tau_3 I_2
\right]
~.
\eeqa

The direction and magnitude 
of linear polarization at selected positions are shown
in Figure \ref{fig:disk} for a stratified disk model with parameters
given in Table \ref{tab:disk pars}, viewed at inclination
$\theta_i=45^\circ$.
Figure \ref{fig:disk}{c,d} show the linear and circular polarization
as a function of azimuthal angle (in the disk plane) for this model.
In addition to accurate results from numerical integration, the results
from the analytic approximation (Eqs. \ref{eq:Q}--\ref{eq:V})
are also plotted.  The analytic
approximation is seen to provide
fair accuracy, even though $\tau_2=0.2$ is not small.

The circular polarization $V/I$ is quite accurate, but 
in Figure \ref{fig:disk}(c), 
the analytic approximation slightly
overestimates the linear polarization fraction.
However, the analytic approximations were developed for $\tau\ll 1$,
and here the total optical depth
$\tau_1+\tau_2+\tau_3=0.3$ is not small.

For this model, the linear polarization varies from $1.4\%$ to $3.2\%$ around
the disk, with average value $\sim$$2.5\%$.
The linear polarization tends to be close to 
the azimuthal direction, with largest
values on the major axis, and smallest values along the minor axis of
inclined disk (see Figure \ref{fig:disk}).

The predicted circular polarization $|V|/I$ 
is small but perhaps detectable, with
$V/I$ varying from positive to negative from one
quadrant to another (see Figure \ref{fig:disk}),
 with maxima $|V|/I \approx 0.2\%$ (see Figure \ref{fig:disk}(d)).

\citet{Stephens+Yang+Li+etal_2017} mapped $V$ over the HL Tau disk
at $3.3\mm$, $1.3\mm$, and $870\micron$.
The $3.3\mm$ $V$ map does not appear to show any statistically significant
detection, with upper limits $|V/I| \ltsim 1\%$.
At $1.3\mm$ and $870\micron$ the NW side of the major axis may have
$V/I \approx -1\%$, but whether this is real rather than an instrumental
artifact remains unclear.  In any event, the likely importance of
scattering at these shorter wavelengths will complicate interpretation.

\section{\label{sec:discussion}
         Discussion}

For typical molecular clouds we conclude that
the circular polarization will
be undetectably
small at the far-infrared and submm wavelengths where the clouds
radiate strongly.  Probing the magnetic field structure in such clouds
using circular polarization is feasible only at shorter infrared wavelengths 
where the extinction is appreciable, using 
embedded infrared sources (stars, protostars, \added{or PDRs}).

The thermal dust emission from
so-called IR dark clouds (IRDCs) in the inner Galaxy -- such
as the ``Brick'' -- can show appreciable levels of linear polarization,
demonstrating both that there is appreciable grain alignment
\emph{and} that the magnetic field structure in the cloud, while showing
evidence of rotation, is relatively coherent.
IRDCs
have large enough column densities that the resulting circular polarization 
may reach detectable levels.  For one position on the Brick and
plausible assumptions concerning the field, 
we estimate a circular polarization
$|V/I| \approx \replaced{0.06}{0.025}\%$ at $850\GHz$.
If the circular polarization can be detected and mapped in IRDCs,
it would provide constraints on the 3-dimensional magnetic field structure.
Unfortunately, the predicted $V/I$ is small, especially
at longer wavelengths (we expect
$V/I \propto \lambda^{-1.1}$), and detection will be
challenging.

Protoplanetary disks may offer the best opportunity to measure
circular polarization at submm wavelengths.
If there are significant
changes in the direction of grain alignment between the dust near the
midplane and dust well above and below
the midplane, linear dichroism and birefringence will
produce circular polarization.
Alignment processes in protoplanetary disks remain uncertain, but we 
suggest that grain drift may cause the grains near the midplane to 
be aligned with long axes
preferentially in the azimuthal direction, while grains above and below
the midplane may
be aligned with long axes
tending to be in the vertical direction (normal to the disk).
If the grains are small enough that scattering can be neglected, we calculate
the linear and circular polarization that would be expected for such
a model.
A characteristic quadrupole pattern of circular polarization is predicted
for this kind of grain alignment (see Figure \ref{fig:disk}).
Eq.\ (\ref{eq:V}) can be used to estimate the circular polarization
at wavelengths $\lambda \gtsim 100\micron$ where thermal emission is
strong and the grains may be approximated by the Rayleigh limit. 
  
We present a simple example to show the linear and circular polarization that
might be present in protoplanetary disks, such as the disk around HL Tau.
This example is not being put forward as a realistic model for HL Tau, but
simply to illustrate the possible circular polarization from dust aligned by
streaming in a stratified disk.
If observed,
this would help clarify the physical processes responsible for grain
alignment in protoplanetary disks.  Absence of this circular polarization
would indicate that the preferred direction for grain
alignment in high-altitude regions is the same as the preferred direction
near the midplane, or else that grain alignment occurs 
only in the midplane, or only in the upper layers.

\added{If circular polarization is detected and mapped in a protoplanetary
disk, 
interpretation will require
radiative transfer 
models that include the birefringence and dichroism discussed here as well
as the circular polarization produced by scattering of linearly polarized
radiation.  
Models will be sensitive to the spatial distribution of the
dust, and also to the sizes and scattering properties of the solid particles.
Maps of $V/I$ at multiple frequencies would strongly constrain 
protoplanetary disk models.}
\section{\label{sec:summary}
         Summary}

\begin{enumerate}
\item 
We present the transfer equations for the Stokes parameters, including
the effects of thermal emission.  
Once the properties of the medium are specified, these equations 
can easily be integrated numerically.
For small optical depths, analytic solutions are given for
clouds with a uniform twist to the magnetic field, and for stratified
clouds with uniform alignment within individual strata.

\item 
Using the ``astrodust'' grain model \citep{Draine+Hensley_2021a}
we calculate the relevant optical properties of dust grains for producing
linear and circular polarization in the far-infrared and submm.
By adjusting the assumed degree of dust alignment $\falign$, these dust
properties may approximate the properties of dust in protoplanetary disks,
at wavelengths where scattering can be neglected.

\item 
At submm wavelengths, the ``phase shift'' cross section $C_{\rm pha}$
tends to be
much larger than the absorption cross section
$C_{\rm abs}$.  
We estimate $C_{\rm pha}/C_{\rm abs} \approx 24(\lambda/\mm)^{0.7}$.

\item 
The far-IR emission from dust in diffuse clouds, and in normal
molecular clouds, will have very low levels of circular polarization,
below current and foreseen sensitivities.

\item 
If the magnetic field in IRDCs has a significant
systematic twist,
the emission from IRDCs \added{ -- such as the ``Brick'' --} may have
$V/I \approx \replaced{0.06}{0.025}\% (\lambda/350\micron)^{-1.1}$

\item 
If dust grains in protoplanetary disks are aligned in different
directions in different strata, the resulting submm emission may
be circularly polarized with peak 
$V/I\approx 0.2\% (\lambda/350\micron)^{-1.1}$
for one simple example
\added{with parameters suggested by HL Tau}.  
Measuring the circular polarization can constrain
the mechanisms responsible for grain alignment in protoplanetary disks.

\end{enumerate}

This work was supported in part by NSF grant AST-1908123.
I thank \added{Yilun Guan,}
Chat Hull and Joseph Weingartner for helpful discussions,
Robert Lupton for availability of the SM package. \added{
I thank the anonymous referee for helpful suggestions that improved 
this paper.}

\bibliography{/u/draine/work/libe/btdrefs}

\appendix
\section{\label{app:uniform twist}
         Uniform Twist}

Assume a single dust temperature $T_d$.
Define $d\tau^\prime \equiv \delta dz$.
Suppose $\Psi$ varies linearly with $\tau$, with total twist $\Delta\Psi$:
\beq \label{eq:linear Psi}
\Psi(\tau^\prime) = \Psi_0 + \alpha\tau^\prime
~~~~,~~~~ \alpha=\frac{\Delta\Psi}{\tau}
~~~.
\eeq
Assuming ${\bf S}=(0,0,0,0)$ for $\tau=0$, and integrating 
Eq.\ (\ref{eq:propagation}) while retaining only low-order 
terms in $\tau$, we obtain:
\beqa \label{eq:I(tau)}
I &~\approx~& B(T_d) \tau\left\{1-\frac{1}{2}\tau - 
\left( \frac{\Delta\sigma}{\delta}\right)^2 \tau
\frac{[1-\cos(2\Delta\Psi)]}{4(\Delta\Psi)^2}
\right\}
\\ \label{eq:Q(tau)}
Q &\approx& -\left(\frac{\Delta\sigma}{\delta}\right)
B(T_d) \tau\left(1-\tau \right)
\frac{
\left[\sin(2\Psi)-\sin(2\Psi_0)\right]}{2\Delta\Psi}
\\ \label{eq:U(tau)}
U &\approx& -\left(\frac{\Delta\sigma}{\delta}\right)
B(T_d) \tau\left(1-\tau\right)
\frac{
\left[\cos(2\Psi_0) - \cos(2\Psi)\right]}{2\Delta\Psi}
\\ \label{eq:V(tau)}
V &\approx& 
\left(\frac{\Delta\sigma}{\delta}\right)
\left(\frac{\Delta\epsilon}{\delta}\right)
\frac{B(T_d) \tau^2}{2\Delta\Psi}
\left\{
1  - \frac{1}{2}\tau
+ \frac{\tau}{(2\Delta\Psi)^2}
  \left[\cos(2\Delta\Psi)-1\right]
- \frac{(1-\tau)}{2\Delta\Psi}\sin(2\Delta\Psi)
\right\}
\\ \label{eq:p}
p &\equiv& \frac{(Q^2+U^2)^{1/2}}{I}
\,\approx\, \frac{1}{\Delta\Psi} 
\left(\frac{\delta\sigma}{\delta}\right)
\frac{(1-\tau)\left[1-\cos(2\Delta\Psi)\right]^{1/2}}
{1 - \frac{1}{2}\tau - \tau\left(\frac{\Delta\sigma}{\delta}\right)^2
\frac{1-\cos(2\Delta\Psi)}{4(\Delta\Psi)^2}}
~~~.
\eeqa 
These results are valid for $\tau\ll 1$, and general twist angle
$\Delta\Psi$.


\added{
\section{\label{app:two zone}
         A Two Zone Model}
Suppose that $\Psi=\Psi_1$ for $0 < \tau < \tau_1$
and $\Psi=\Psi_1+ \Delta\Psi(\tau-\tau_1)/\tau_2$ 
for $\tau_1 < \tau < \tau_1+\tau_2$.
Let $T_d=T_{d1}$ for $0 < \tau < \tau_1$, and
$T_d=T_{d2}$ for $\tau_1 < \tau < \tau_1+\tau_2$.
Define
\beqa
A &\equiv& \frac{\Delta\sigma}{\delta}\left[I_1-B(T_{d2})\right]
~~~.
\eeqa
Assuming ${\bf S}=(0,0,0,0)$ for $\tau=0$, integrating Eq.\ (\ref{eq:propagation}),
retaining only low-order terms in $\tau$, we obtain 
${\bf S}_1=(I_1,Q_1,U_1,V_1)$
at $\tau=\tau_1$,
and ${\bf S}_2=(I_2,Q_2,U_2,V_2)$ at $\tau=\tau_1+\tau_2$:
\beqa
I_1 &=& B(T_{d1})(1-e^{-\tau_1})
\\
Q_1 &=& - \left(\frac{\Delta\sigma}{\delta}\right)_1 \cos(2\Psi_1) I_1
\\
U_1 &=& - \left(\frac{\Delta\sigma}{\delta}\right)_1 \sin(2\Psi_1) I_1
\\
V_1 &=& 0
\\
I_2 &\approx& I_1 e^{-\tau_2} + B(T_{d2})(1-e^{-\tau_2})
\\
Q_2 &\approx& 
Q_1(1-\tau_2) + A\tau_2(1-\tau_2)\frac{\sin(2\Psi_2)-\sin(2\Psi_1)}{2\Delta\Psi}
\\
U_2 &\approx& 
U_1(1-\tau_2) + A\tau_2(1-\tau_2)\frac{\cos(2\Psi_1)-\sin(2\Psi_2)}{2\Delta\Psi}
\eeqa
\beqa
V_2 &\approx& 
\left(\frac{\Delta\epsilon}{\delta}\right)_{\!\!2}
 Q_1
\left[\tau_2
\frac{\cos2\Psi_2-\cos2\Psi_1}{2\Delta\Psi}-
\tau_2^2 \frac{\sin2\Psi_2-\sin2\Psi_1-2\Delta\Psi\cos2\Psi_2}{2\Delta\Psi}
\right]
\\ \nonumber
&&+
\left(\frac{\Delta\epsilon}{\delta}\right)_{\!\!2}
 U_1
\left[\tau_2
\frac{\sin2\Psi_2-\sin2\Psi_1}{2\Delta\Psi}-
\tau_2^2 \frac{\cos2\Psi_2-\cos2\Psi_1-2\Delta\Psi\sin2\Psi_2}{2\Delta\Psi}
\right]
\\ \nonumber
&&
-\left(\frac{\Delta\epsilon}{\delta}\right)_{\!\!2} 
A \frac{\tau_2}{2\Delta\Psi}\left[\tau_2-\frac{1}{2}\tau_2^2\right]
+\left(\frac{\Delta\epsilon}{\delta}\right)_{\!\!2}
A \left(\frac{\tau_2}{2\Delta\Psi}\right)^2\sin2\Delta\Psi
\\ \nonumber
&& 
+\left(\frac{\Delta\epsilon}{\delta}\right)_{\!\!2}
A \left(\frac{\tau_2}{2\Delta\Psi}\right)^3
\left[1-\cos2\Delta\Psi - 2\Delta\Psi\sin2\Delta\Psi\right] + O(\tau_2^4)
~~~.
\eeqa
}

\section{\label{app:three zone model}Three Zone Model}

Suppose the dust is located in three zones, 
with dust temperatures $T_{d1}$, $T_{d2}$, and $T_{d3}$.
The aligned dust grains have $\Psi=\Psi_1$ for
$0 < \tau < \tau_1$, $\Psi=\Psi_2$ for 
$\tau_1 < \tau < \tau_1+\tau_2$, and
$\Psi=\Psi_3$ for 
$\tau_1+\tau_2 < \tau < \tau_1+\tau_2+\tau_3$.
Suppose all $\tau_{j} \ll 1$.

Define
\beqa
I_1&\,\equiv\,& 
B(T_{d1}\left[1-e^{-\tau_1}\right]e^{-\tau_2-\tau_3}
\,\approx\,
B(T_{d1})\, \tau_1\left[1-\frac{1}{2}\tau_1\right]e^{-\tau_2-\tau_3}
\\
I_2&\equiv& 
B(T_{d2})\left[1-e^{-\tau_2}\right]e^{-\tau_3}
\,\approx\,
B(T_{d2})\, \tau_2\left[1-\frac{1}{2}\tau_2\right]e^{-\tau_3}
\\
I_3&\equiv& 
B(T_{d3})\left[1-e^{-\tau_3}\right]
\,\approx\,
B(T_{d3})\, \tau_3\left[1-\frac{1}{2}\tau_3\right]
\eeqa

If ${\bf S}=(0,0,0,0)$ for $\tau=0$, then the radiation emerging from layer 3
has
\beqa
I &\,\approx\,& I_1+I_2+I_3
\\ 
Q &\approx&
-\left(\frac{\Delta\sigma}{\delta}\right)_{\!1}\!\cos(2\Psi_1)
I_1
-\left(\frac{\Delta\sigma}{\delta}\right)_{\!2}\!\cos(2\Psi_2)
[I_2-\tau_2 I_1]
-\left(\frac{\Delta\sigma}{\delta}\right)_{\!3}\!\cos(2\Psi_3)
[I_3-\tau_3(I_1+I_2)]~~~~~
\\
U &\approx&
-\left(\frac{\Delta\sigma}{\delta}\right)_{\!1}\!\sin(2\Psi_1)
I_1
-\left(\frac{\Delta\sigma}{\delta}\right)_{\!2}\!\sin(2\Psi_2)
[I_2-\tau_2I_1]
-\left(\frac{\Delta\sigma}{\delta}\right)_{\!3}\!\sin(2\Psi_3)
[I_3-\tau_3(I_1+I_2)]~~~~~
\\ \nonumber
V & \approx & 
\left(\frac{\Delta\epsilon}{\delta}\right)_{\!2}
\left(\frac{\Delta\sigma}{\delta}\right)_{\!1}
\sin\left(2\Psi_2-2\Psi_1\right)
\tau_2 I_1
+ 
\left(\frac{\Delta\epsilon}{\delta}\right)_{\!3}
\left(\frac{\Delta\sigma}{\delta}\right)_{\!1}
\sin\left(2\Psi_3-2\Psi_1\right)
\tau_3 I_1
\\
&&
+ 
\left(\frac{\Delta\epsilon}{\delta}\right)_{\!3}
\left(\frac{\Delta\sigma}{\delta}\right)_{\!2}
\sin\left(2\Psi_3-2\Psi_2\right)
\tau_3 I_2
~~.
\eeqa

\end{document}